\documentclass[journal=jacsat,manuscript=article]{achemso}

\usepackage[version=3]{mhchem} % Formula subscripts using \ce{}

\author{Jinkyung Kim}
\altaffiliation{These authors contributed equally to this work.}
\affiliation[QNS]
{Center for Quantum Nanoscience (QNS), Institute for Basic Science (IBS), Seoul 03760, South Korea}
\alsoaffiliation[Department of Physics]
{Department of Physics, Ewha Womans University, Seoul 03760, South Korea}

\author{Kyungju Noh}
\altaffiliation{These authors contributed equally to this work.}
\affiliation[QNS]
{Center for Quantum Nanoscience (QNS), Institute for Basic Science (IBS), Seoul 03760, South Korea}
\alsoaffiliation[Department of Physics]
{Department of Physics, Ewha Womans University, Seoul 03760, South Korea}

\author{Yi Chen}
\altaffiliation{These authors contributed equally to this work.}
\affiliation[QNS]
{Center for Quantum Nanoscience (QNS), Institute for Basic Science (IBS), Seoul 03760, South Korea}
\alsoaffiliation[Ewha]
{Ewha Womans University, Seoul 03760, Republic of Korea}

\author{Fabio Donati}
\affiliation[QNS]
{Center for Quantum Nanoscience (QNS), Institute for Basic Science (IBS), Seoul 03760, South Korea}
\alsoaffiliation[Department of Physics]
{Department of Physics, Ewha Womans University, Seoul 03760, South Korea}

\author{Andreas J. Heinrich}
\email{heinrich.andreas@qns.science}
\affiliation[QNS]
{Center for Quantum Nanoscience (QNS), Institute for Basic Science (IBS), Seoul 03760, South Korea}
\alsoaffiliation[Department of Physics]
{Department of Physics, Ewha Womans University, Seoul 03760, South Korea}

\author{Christoph Wolf}
\email{wolf.christoph@qns.science}
\affiliation[QNS]
{Center for Quantum Nanoscience (QNS), Institute for Basic Science (IBS), Seoul 03760, South Korea}
\alsoaffiliation[Ewha]
{Ewha Womans University, Seoul 03760, Republic of Korea}

\author{Yujeong Bae}
\email{bae.yujeong@qns.science}
\affiliation[QNS]
{Center for Quantum Nanoscience (QNS), Institute for Basic Science (IBS), Seoul 03760, South Korea}
\alsoaffiliation[Department of Physics]
{Department of Physics, Ewha Womans University, Seoul 03760, South Korea}

\title[An \textsf{achemso} demo]
{Anisotropic hyperfine interaction of surface-adsorbed single atoms}
\abbreviations{IR,NMR,UV}
\keywords{Scanning tunneling microscopy, Electron spin resonance, Vector field, Hyperfine interaction, Nuclear spin \LaTeX}

\begin{document}

% \begin{tocentry}

% % Any ideas on how to add TOC here?

% % \begin{figure}
% %   \includegraphics[width=0.7 \textwidth ]{Figures/ToC.png}
% %   \caption*
% %   \label{ToC}
% % \end{figure}

% \end{tocentry}

\begin{abstract}
  Hyperfine interactions between electron and nuclear spins have been widely used in material science, organic chemistry, and structural biology as a sensitive probe to the local chemical environment through spatial identification of nuclear spins. With the nuclear spins identified, the isotropic and anisotropic components of the hyperfine interactions in turn offer unique insight into the electronic ground-state properties of the paramagnetic centers. However, traditional ensemble measurements of hyperfine interactions average over a macroscopic number of spins with different geometrical locations and nuclear isotopes. Here, we use a scanning tunneling microscope (STM) combined with electron spin resonance (ESR) to measure hyperfine spectra of hydrogenated-titanium (Ti) atoms on MgO/Ag(100) and thereby determine the isotropic and anisotropic hyperfine interactions at the single-atom level. By combining vector-field ESR spectroscopy with STM-based atom manipulation, we characterize the full hyperfine tensor of individual $^{47}$Ti and $^{49}$Ti atoms and identify significant spatial anisotropy of hyperfine interaction for both isotopes when they are adsorbed at low-symmetry binding sites. Density functional theory calculations reveal that the large hyperfine anisotropy arises from a highly anisotropic distribution of the ground-state electron spin density. Our work highlights the power of ESR-STM-enabled single-atom hyperfine spectroscopy as a powerful tool in revealing ground-state electronic structures and atomic-scale chemical environments with nano-electronvolt resolution.

  %Hyperfine interactions between electron and nuclear spins provide sensitive probes to the chemical environment and electronic structures of atoms, molecules, and crystal defects. Also, the growing interest in using hyperfine interactions as an alternative way of controlling single nuclear spins stimulates simultaneous investigations on the hyperfine interaction and the atomic scale environment, which necessitates high spectral and spatial resolutions. Here, we precisely determine the isotropic and anisotropic hyperfine interactions of a single atom, hydrogenated-titanium, and, at the same time, its local environment on MgO/Ag(100) using electron spin resonance in a scanning tunneling microscope. By means of atom manipulation in a vector magnetic field, we observe a significantly large anisotropy of hyperfine interactions along the three principal axes for two isotopes of Ti. With subtle contribution of the anisotropy of g-factors, the hyperfine interaction reflects the anisotropic distribution of electron spin density in the electronic ground state. Our work highlights the strength of the single spin hyperfine spectroscopy as a probe to the electronic structure and chemical environment at atomic scale.
  
\end{abstract}


%Intro: why hyperfine splitting is important to be investigated at atomic scale.

\begin{figure}
  \includegraphics[width=0.5 \textwidth ]{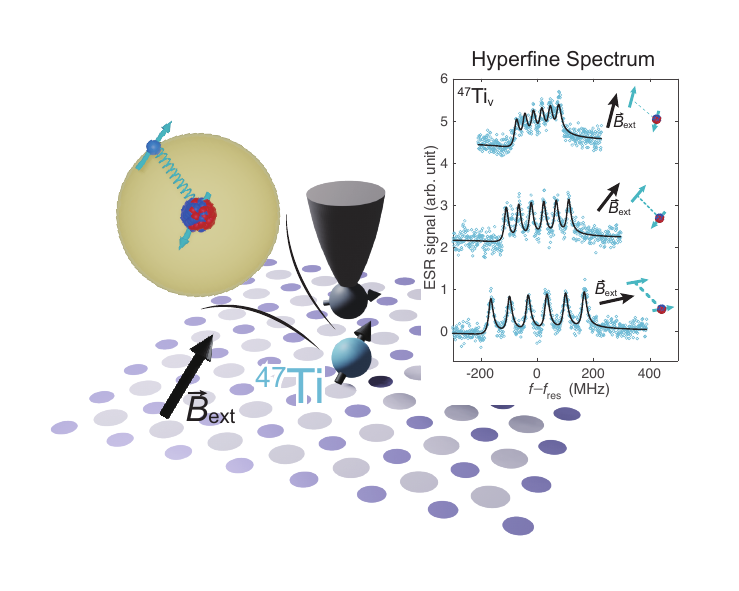}
  \caption*{Keywords: Scanning tunneling microscopy, Electron spin resonance, Scanning probe, Hyperfine interaction, Nuclear spin}
\end{figure}

Conventional ensemble magnetic resonance techniques have been extensively used for probing hyperfine interactions between paramagnetic spin centers and nearby nuclear spins, where the sensitivity largely depends on the spin concentration\cite{nla.cat-vn1762031}. Hyperfine interactions at the single spin level have recently attracted significant interests due to promise in sensitive detection of the local chemical environment \cite{doi:10.1126/science.aal2538} and nuclear-spin-based quantum information processing \cite{Kane:1998wh,PhysRevX.4.021044,Abobeih2019,Zhao2012}. These new scientific endeavors are enabled by technological developments that allow electron spin resonance (ESR) operations at the single-spin level\cite{Rugar2004}, for example, through optically addressable color centers in insulators\cite{Doherty2013,Schirhagl2014} or semiconductor donor atoms equipped with nanofabricated charge detectors \cite{RevModPhys.85.961}. However, for a general paramagnetic center placed in its native chemical environment, characterization of the hyperfine interactions at the single-spin level has been extremely difficult.

Scanning tunneling microscopy (STM) with ESR capabilities offers a new appealing platform for \textit{in-situ} characterization of individual spin-carrying atoms and molecules \cite{Baumann417,Zhang2022,https://doi.org/10.1002/adma.202107534}. ESR-STM spectroscopy offers tens of nano-electrovolt energy resolution, far beyond traditional STM bias spectroscopy\cite{PhysRev.165.821, Song2010, PhysRevLett.107.076804}, thus allowing for probing hyperfine interactions at the atomic scale\cite{Willke336,Yang2018}. When combined with STM's single-atom selectivity, hyperfine interactions from single atoms with different isotopes and different binding sites can be individually determined without any spatial averaging\cite{Koehler1995}.

Here we use a state-of-the-art ESR-STM system to measure the full hyperfine tensor of single hydrogenated $^{47}$Ti and $^{49}$Ti atoms on MgO/Ag(100). Using a vector magnetic field and STM-based atom manipulation, we quantify the isotropic and anisotropic hyperfine interactions for the two Ti isotopes. A large hyperfine anisotropy of more than 67$\%$ is observed for both Ti isotopes on a low-symmetry binding site, which reveals a highly anisotropic distribution of the ground-state spin density that is consistent with density functional theory (DFT) results.

\begin{figure}
  \includegraphics[width=0.7 \textwidth ]{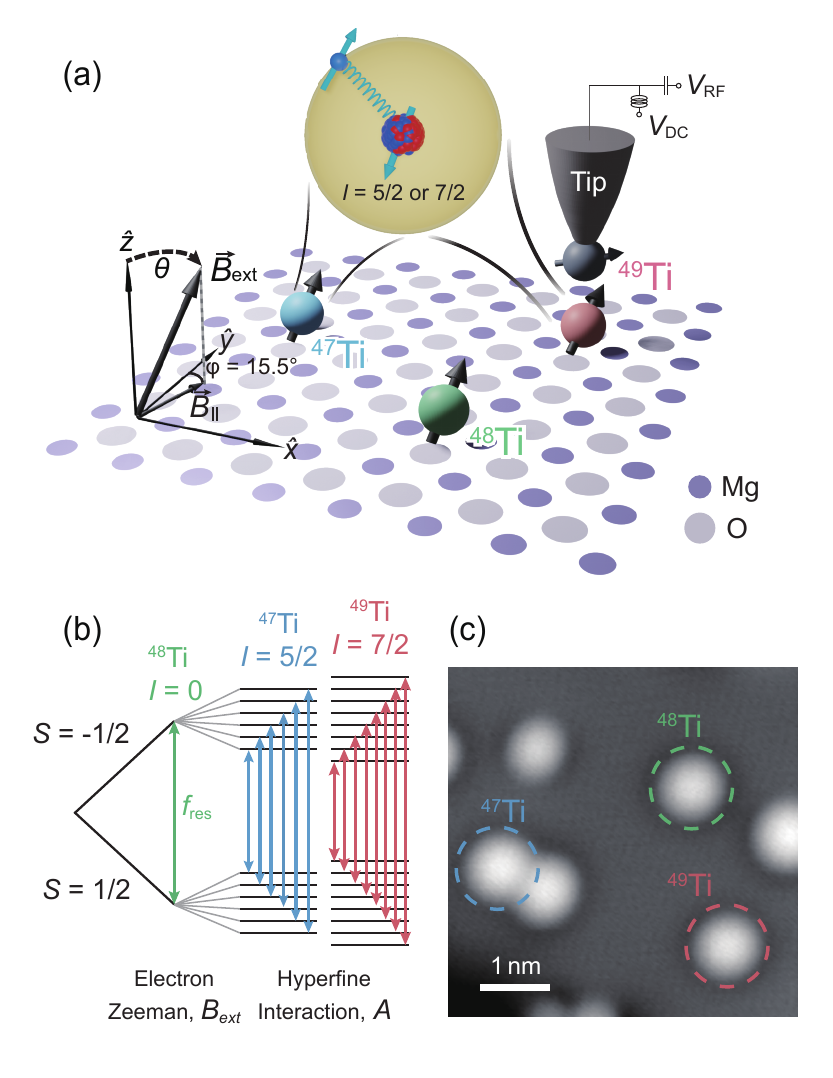}
  \caption{Hyperfine interaction of single hydrogenated titanium (Ti) atoms on the MgO/Ag(100) surface. (a) Schematic of the ESR-STM measurements on different isotopes of Ti under a rotatable magnetic field. Single-atom ESR spectroscopy is performed by detecting the change of spin-polarized tunnel current under resonant driving. The MgO lattice directions are indicated by $\hat{x}$ and $\hat{y}$, whereas $\hat{z}$ is the out-of-plane direction. A vector external magnetic field $\boldsymbol{B}_{\mathrm{ext}}$ is applied in a plane which is 15.5$^{\circ}$ rotated from the $yz$-plane around the $\hat{z}$-axis. The rotation angle of $\boldsymbol{B}_{\mathrm{ext}}$ relative to the out-of-plane direction is labelled as $\theta$. (b) Schematic of energy levels and ESR transitions of different Ti isotopes in the presence of a magnetic field $B_\mathrm{ext}$ and the hyperfine interaction $A$.
  %\textsuperscript{47}Ti and \textsuperscript{49}Ti with nonzero nuclear spin ($I=5/2$ and $I=7/2$), compared to that of \textsuperscript{48}Ti with zero nuclear spin 0, in the presence of the magnetic field $B$ and the hyperfine interaction $A$. 
  The ESR transitions are denoted by double sided arrows. (c) STM image showing three different types of Ti isotopes. Different Ti isotopes show identical topographic features but can be readily identified through ESR spectroscopy (Figure 2 and S1) ($V_{\mathrm{DC}}= 100$ mV, $I=20$ pA).}
  \label{figure1}
\end{figure}

%Experimental set-up
We performed ESR-STM experiments using a home-built STM system equipped with RF cabling and a two-axis vector magnet \cite{PhysRevB.104.174408}. The vector magnet provides an in-plane magnetic field up to 9 T and an out-of-plane field up to 2 T. Individual Ti atoms were deposited on two monolayers of MgO(100) grown on a Ag(100) substrate while the sample was kept below 10 K in the STM stage (Figure \ref{figure1}a). Evaporation of Ti was performed using a commercial electron-beam evaporator. Ti rods with natural isotope abundance were used, in which the most abundant isotope, \textsuperscript{48}Ti, has zero nuclear spin, while \textsuperscript{47}Ti (with 7.4\% abundance) has a nonzero nuclear spin of $I = 5/2$ and \textsuperscript{49}Ti (with 5.4\% abundance) has a nuclear spin of $I = 7/2$ (Figure \ref{figure1}b) \cite{haynes2016crc}. Due to the residual hydrogen gas in the vacuum chamber,\cite{Natterer2013} Ti atoms deposited on the MgO surface are most likely hydrogenated and are known to have electron spin-1/2.\cite{PhysRevLett.119.227206,Baeeaau4159} Here, we denote the hydrogenated Ti atoms as Ti for simplicity. Ti atoms were found at two different types of binding sites on MgO(100), atop of an oxygen atom \cite{PhysRevLett.119.227206,Steinbrecher:2018ux} or at a bridge site between two oxygen atoms \cite{Baeeaau4159,Seiferteabc5511}. In this work, we focus on Ti atoms at the bridge binding sites, which have a significantly larger hyperfine coupling than the oxygen binding sites \cite{Willke336}. Note that in our experimental set-up, the in-plane magnetic-field axis is 15.5$^{\circ}$ tilted from the oxygen lattice direction of MgO(100) (Figure \ref{figure1}a). This results in two non-identical bridge binding sites of Ti, referred to as Ti near the vertical direction (Ti\textsubscript{v}) and Ti near the horizontal direction (Ti\textsubscript{h}) \cite{PhysRevB.104.174408,Veldman964}. These two inequivalent in-plane sites, in combination with the two-axis magnetic field, allow us to determine the hyperfine interaction tensor as discussed below.

\begin{figure}
  \includegraphics[width=1 \textwidth ]{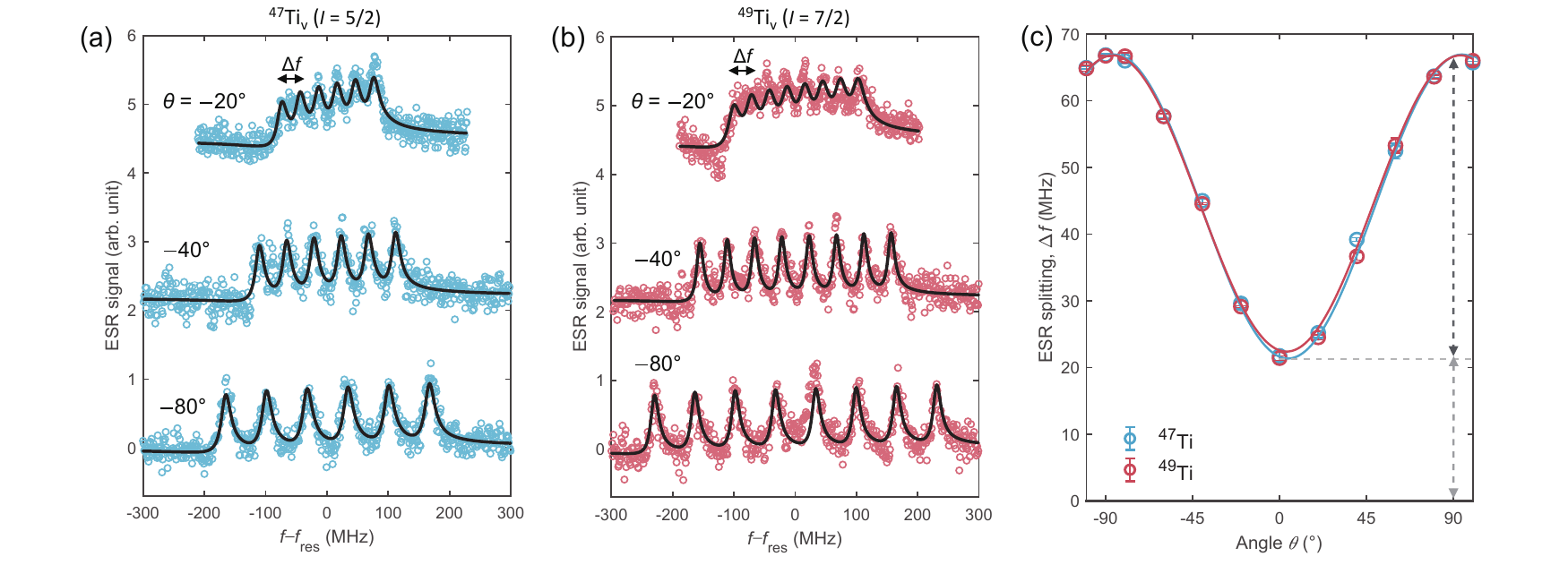}
  \caption{Angular variations of hyperfine interactions of \textsuperscript{47}Ti\textsubscript{v} and \textsuperscript{49}Ti\textsubscript{v}. (a) Hyperfine spectra of \textsuperscript{47}Ti\textsubscript{v} and (b) \textsuperscript{49}Ti\textsubscript{v} measured at different angles $\theta$ of the external magnetic field. For both isotopes, the hyperfine splitting is greatly increased when the field is rotated away from the out-of-plane direction (i.e., for the shown data, towards more negative $\theta$). Fits in (a) yield a hyperfine splitting of 30.0\textpm0.4 MHz at $\theta = -20^{\circ}$, 44.4\textpm0.3 MHz at $\theta = -40^{\circ}$, and 66.5\textpm0.2 MHz at $\theta = -80^{\circ}$. In (b), the hyperfine splitting is determined to be 29.0\textpm0.3 MHz at $\theta = -20^{\circ}$, 44.6\textpm0.1 MHz at $\theta = -40^{\circ}$, and 65.8\textpm0.2 MHz at $\theta = -80^{\circ}$. ESR spectra are plotted against $f - f_{0}$, where $f_0$ is the resonance frequency of a \textsuperscript{48}Ti\textsubscript{v} atom (with zero nuclear spin) measured under the same conditions. ESR spectra are normalized in intensity, and successive curves are vertically shifted for clarity ($V$\textsubscript{DC} = 40 mV, $I$\textsubscript{set} = 1.5$\sim$8 pA, $V$\textsubscript{RF} = 15$\sim$50 mV). (c) Hyperfine splittings of \textsuperscript{47}Ti\textsubscript{v} and \textsuperscript{49}Ti\textsubscript{v} measured as a function of the magnetic field angle, $\theta$. Solid curves are guides to the eye. Dashed arrows highlight the significant hyperfine anisotropy between in-plane and out-of-plane directions. 
  %The solid curves are the fit to Eq. \ref{eq: B-orientation dependent hyperfine}. 
  Error bars are given by the standard errors of different measurements under the same conditions. From these measurements we conclude that there is no noticeable difference in the hyperfine splittings of \textsuperscript{47}Ti and \textsuperscript{49}Ti.}  
  \label{figure2}
\end{figure}

%Introduce isotopes
Figure \ref{figure1}c shows a typical STM image containing three different types of Ti isotopes. While Ti atoms with different isotopes show identical STM topographic and bias spectroscopic features, they are clearly distinguishable from STM-based ESR spectroscopy (Figure S1). Out of 94 Ti atoms that we measured, the majority (83 atoms) show only one ESR peak, corresponding to the most abundant \textsuperscript{48}Ti isotope with zero nuclear spin (Figure \ref{figure1}b and S1). 6 Ti atoms (roughly 6.4\%) exhibit 6 ESR peaks that correspond to \textsuperscript{47}Ti with $I = 5/2$, and 5 Ti atoms (roughly 5.3\%) exhibit 8 ESR peaks that correspond to \textsuperscript{49}Ti with $I = 7/2$ (Figure \ref{figure1}b and S1). The hyperfine splitting is much smaller than the thermal energy at 0.6 K, resulting in nearly equal populations in the nuclear spin states and thus equal peak intensities. The ESR peaks observed in our measurements are found to be equally spaced (see, e.g., Figure \ref{figure2} and S1), which indicates negligible contributions of the nuclear Zeeman interaction and the electric quadrupole interaction\cite{abragam2012electron}. These considerations lead us to write down a simplified spin Hamiltonian of the \textsuperscript{47}Ti and \textsuperscript{49}Ti atoms as
%In the parameter regime used in our measurements, the magnitude of the external magnetic field (0.8 T) results in an electronic Zeeman energy is much larger than the hyperfine interaction and the nuclear Zeeman energy for Ti ($\gamma_{n} \sim \gamma_{e} \cdot 8.6 \times 10^{-5}$ for the gyromagnetic ratios of electron spin, $\gamma_{e}$, and Ti nuclear spin, $\gamma_{n}$) is negligibly small. The Hamiltonian for a single Ti atom with a nonzero nuclear spin reads
\begin{equation}
\boldsymbol{\mathcal{H}} = \mu_\mathrm{B} \boldsymbol{B}_{\mathrm{ext}} \cdot \mathbf{g} \cdot \boldsymbol{S} + \boldsymbol{S} \cdot \mathbf{A} \cdot \boldsymbol{I} =  \mu_\mathrm{B} \boldsymbol{B}_{\mathrm{ext}} \cdot \mathbf{g} \cdot \boldsymbol{S} + A_{\mathrm{iso}}\boldsymbol{S} \cdot \boldsymbol{I} + \boldsymbol{S} \cdot \mathbf{T} \cdot \boldsymbol{I},%+\mathcal{O}\left(\boldsymbol{I}^2\right),
    \label{eq: spin-Hamiltonian}
\end{equation}
where $\boldsymbol{S}$ and $\boldsymbol{I}$ are the electron and nuclear spin operators, respectively, $\boldsymbol{B}_{\mathrm{ext}}$ is the external magnetic field, and $\mathbf{g}$ is the electron $g$-tensor. $\mathbf{A}$ is the tensor of hyperfine interaction that can be decomposed into an isotropic contact term, $A_{\mathrm{iso}}$, that originates from the direct overlap between the electronic wavefunction and the nuclear spin, and an anisotropic contribution, $\mathbf{T}$, that originates from dipolar interactions between the electron and nuclear spins\cite{abragam2012electron}.
%The isotropic (or Fermi contact) contribution, $A_{\mathrm{iso}}$, is determined by the amplitude of electronic wave function at the nucleus and thus the $s$ orbital contribution to the electronic ground state. The anisotropic dipolar interaction, noted as a tensor $\mathbf{T}$, provides a measure of $d$ and/or $p$ orbital character of the unpaired electron and is the source of the anisotropy in the hyperfine interaction.
The principal axes for $\mathbf{g}$ and $\mathbf{A}$ tensors coincide with the crystalline lattice axes of MgO to match the $C_{2v}$ symmetry of Ti on bridge sites\cite{abragam2012electron}. %$\mathcal{O}\left(I^2\right)$ indicates higher order terms, most notably the nuclear-electron quadrupole interaction, which, as we will show, does not play a role for Ti adsorbed at a bridge site.

Different components of the $\mathbf{A}$ tensor can be probed by changing the direction of the external magnetic field $\boldsymbol{B}_{\mathrm{ext}}$. Under the external magnetic field $\boldsymbol{B}_{\mathrm{ext}}$ of 0.8 T (which is used in all measurements shown in the main text), the electron Zeeman term of the Hamiltonian in eq. \ref{eq: spin-Hamiltonian} dominates and determines the electron spin direction to be almost parallel with $\boldsymbol{B}_{\mathrm{ext}}$ (some deviation arises from $\mathbf{g}$-factor anisotropy \cite{PhysRevB.104.174408}). The electron spin direction in turn determines the nuclear spin direction through the hyperfine coupling $\mathbf{A}$. The rotation of $\boldsymbol{B}_{\mathrm{ext}}$ thus rotates the electron spin $\boldsymbol{S}$ and the nuclear spin $\boldsymbol{I}$ together with it, allowing us to probe different components of $\mathbf{A}$ through the term $\boldsymbol{S} \cdot \mathbf{A} \cdot \boldsymbol{I}$. The energy change associated with the hyperfine interaction, $\boldsymbol{S} \cdot \mathbf{A} \cdot \boldsymbol{I}$, is detected through the splitting between adjacent ESR peaks, $\Delta f$. % because the ESR frequencies between adjacent peaks differ by $\Delta f = \Delta m_{S} \Delta m_{I} A(\theta)$ with $\Delta m_{S}=1$ and $\Delta m_{I}=1$. 
For a Ti atom with $S=1/2$, the ESR transition associated with a certain nuclear spin state, $m_I$, has a frequency of $hf_{m_I}=\mu_{\mathrm{B}} B_{\mathrm{ext}} g(\theta)+A(\theta)m_I$, where $h$ is the Planck's constant and $g(\theta)$ and $A(\theta)$ are experimentally probed $g$-factor and hyperfine constant, respectively, at the field angle $\theta$. The frequency splitting between adjacent ESR lines thus directly yields the hyperfine interaction, since $\Delta f = \lvert f_{m_I}-f_{m_I \pm 1} \rvert = A(\theta)$. The relation between experimentally probed $A(\theta)$ and the principal values of the $\mathbf{A}$ tensor will be discussed later (see eq. \ref{eq: binding site hyperfine}).
%For the adjacent nuclear spin state $m_I \pm 1$, the ESR transition appears at the frequency $hf_{I \pm 1}=\mu_{\mathrm{B}} B_{\mathrm{ext}} g(\theta)+A(\theta)(I \pm 1)$. Thus, the splitting between adjacent ESR peaks is given by $\Delta f = \lvert f_{I}-f_{I \pm 1} \rvert = A(\theta)$. Here $g(\theta)$ and $A(\theta)$ are experimentally probed g-factor and hyperfine constant, respectively, at a magnetic field angle $\theta$. The relation between $A(\theta)$ and the principal values of the $\mathbf{A}$ tensor will be discussed later (see eq. \ref{eq: binding site hyperfine}).

%ESR for isotopes
%Ti isotopes with nonzero nuclear spin can be identified readily from ESR spectroscopy. 

Figure \ref{figure2} shows the experimental results at different field angles. At $\theta=-20^{\circ}$ (close to the out-of-plane direction, see Figure \ref{figure1}a), the ESR splittings for \textsuperscript{47}Ti\textsubscript{v} (Figure \ref{figure2}a) and \textsuperscript{49}Ti\textsubscript{v} (Figure \ref{figure2}b) are measured to be 30.0\textpm0.4 MHz and 29.0\textpm0.3 MHz, respectively. We find that the energy splittings $\Delta f$ measured for \textsuperscript{47}Ti\textsubscript{v} and \textsuperscript{49}Ti\textsubscript{v} are equal within the uncertainty of our measurements, as expected from their identical electronic ground states and very similar nuclear magnetic moments \cite{Mabbs}. The hyperfine interaction is significantly increased when we rotate the magnetic field closer to an in-plane direction as shown in Figure \ref{figure2}a, b. %In addition we observed a shift of the ESR resonance frequencies (Supporting Information Sec.5) which is caused by the anisotropy of the g-value. \cite{PhysRevB.104.174408} 
The large angular variations of the hyperfine splittings for \textsuperscript{47}Ti\textsubscript{v} and \textsuperscript{49}Ti\textsubscript{v} are summarized in Figure \ref{figure2}c. In both cases, the largest splitting of $\sim$67 MHz is observed at $\theta \approx 90^{\circ}$ when $\boldsymbol{B}_\mathrm{ext}$ is applied along an in-plane direction, while the smallest splitting of $\sim$22 MHz is obtained along the out-of-plane direction. These observations indicate very large hyperfine anisotropy (around 67$\%$, see Figure \ref{figure2}c) for both \textsuperscript{47}Ti\textsubscript{v} and \textsuperscript{49}Ti\textsubscript{v} atoms. Due to their nearly identical hyperfine interactions, we will focus on \textsuperscript{47}Ti in the following.

\begin{figure}
  \includegraphics[width=1 \textwidth ]{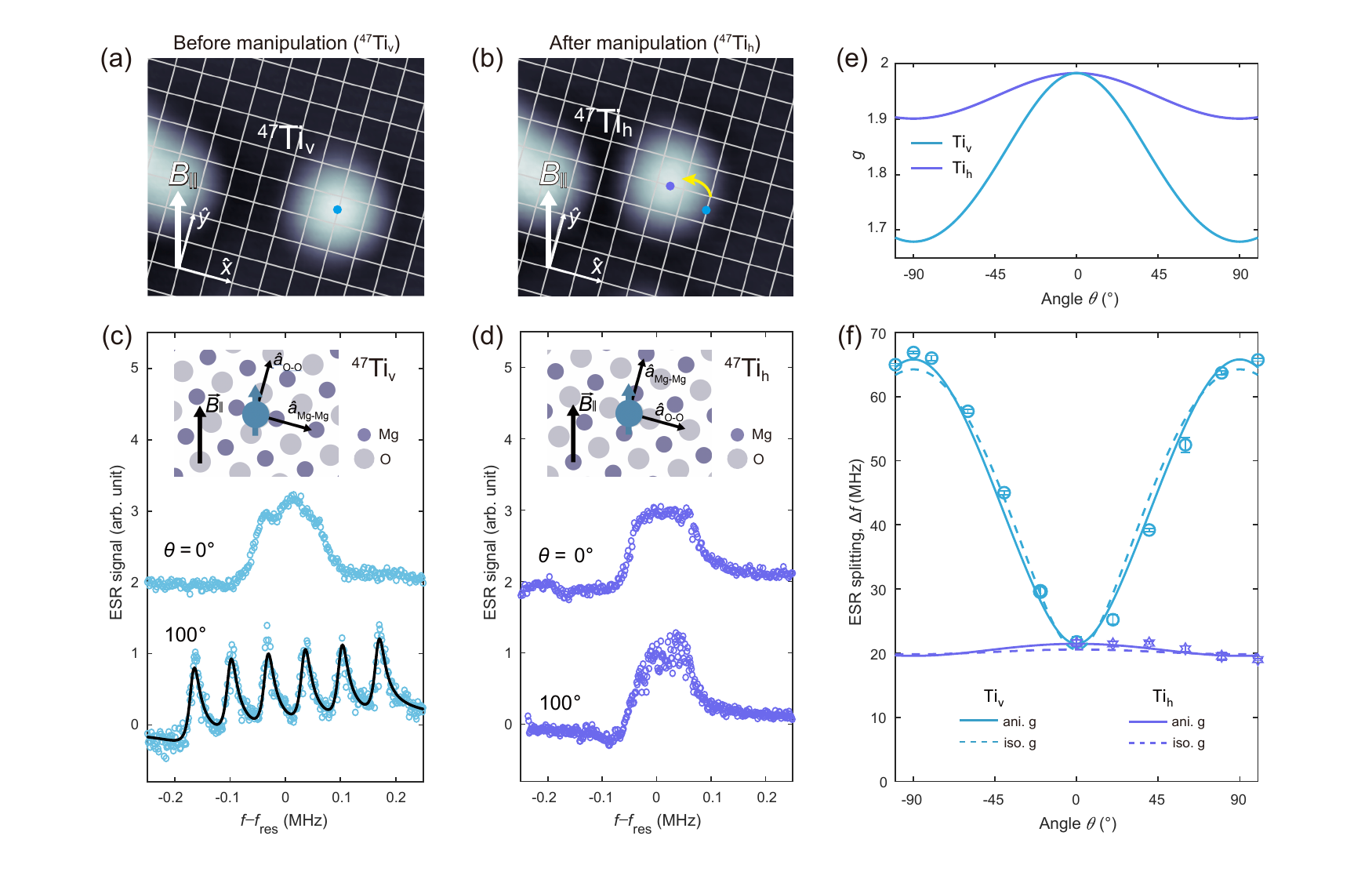}
  \caption{Binding-site-dependent hyperfine spectra of Ti atoms on MgO. (a) STM image of \textsuperscript{47}Ti at a vertical bridge binding site (\textsuperscript{47}Ti\textsubscript{v}). The intersection of grid lines corresponds to the location of oxygen atoms in the MgO lattice. (b) STM image of \textsuperscript{47}Ti\textsubscript{h} taken after moving \textsuperscript{47}Ti\textsubscript{v} in (a) by 1.5$\times$0.5 oxygen lattices, to the horizontal site. The white arrow in each STM image indicates the in-plane magnetic field direction ($\boldsymbol{B}_{||}$) with respect to the MgO lattice (scan conditions: $V_{\mathrm{DC}}= 100$ mV, $I=20$ pA). (c) Hyperfine spectra of \textsuperscript{47}Ti\textsubscript{v} and (d) \textsuperscript{47}Ti\textsubscript{h}, measured at $\theta=0^{\circ}$ (upper) and $100^{\circ}$ (lower). The curves are normalized to unity. Successive curves are vertically shifted for clarity ($V_{\mathrm{DC}}=40$ mV, $I=12\sim20$ pA, $V_{\mathrm{RF}}=15\sim20$ mV). Insets show that $\boldsymbol{B}_{||}$ is close to the O-O direction for \textsuperscript{47}Ti\textsubscript{v} and Mg-Mg direction for \textsuperscript{47}Ti\textsubscript{h}. (e) $g$-factors of bridge-site Ti as a function of the angle $\theta$ as calculated based on the anisotropic $g$-factors determined in ref. \cite{PhysRevB.104.174408}. (f) Anisotropic hyperfine coupling of \textsuperscript{47}Ti\textsubscript{v} and \textsuperscript{47}Ti\textsubscript{h} measured at different magnetic field angles. The O-O direction, close to the in-plane direction of \textsuperscript{47}Ti\textsubscript{v}, shows a significantly higher hyperfine interaction strength than the other two principal axes. Solid and dashed curves correspond to fits to eq. \ref{eq: binding site hyperfine} with anisotropic and isotropic g-values, respectively. The hyperfine splittings at small field angles in Figure \ref{figure3}f are obtained using a different STM tip that allows better resolution (see Figure S3).}
  \label{figure3}
\end{figure}

%Figure3
To determine the hyperfine interaction along the third axis that is not in the tunable plane of the external magnetic field (i.e., the $x$ axis in Figure \ref{figure1}a), we exploit two inequivalent bridge binding sites of Ti on MgO (Ti\textsubscript{v} and Ti\textsubscript{h}). Using atom manipulation, Ti atoms can be moved reversibly between different binding sites on MgO\cite{PhysRevLett.119.227206}. Figure \ref{figure3}a, b shows STM images of \textsuperscript{47}Ti\textsubscript{v} and \textsuperscript{47}Ti\textsubscript{h} taken before and after a typical atom manipulation. Strikingly, this manipulation significantly changes hyperfine spectra as shown in Figure \ref{figure3}c, d. With $\boldsymbol{B}_{\mathrm{ext}}$ applied along the out-of-plane direction (at $\theta = 0 ^\circ$), Ti\textsubscript{v} and Ti\textsubscript{h} are at physically identical sites and exhibit very similar hyperfine spectra (upper curves in Figure \ref{figure3}c, d). With $\boldsymbol{B}_{\mathrm{ext}}$ applied along the in-plane direction, however, the large anisotropy observed in Ti\textsubscript{v} is almost absent in Ti\textsubscript{h} (lower curves in Figure \ref{figure3}c, d). This trend is clearly shown in Figure \ref{figure3}f where a large change of the hyperfine splitting is only observed for Ti\textsubscript{v}. Since the in-plane field direction of Ti\textsubscript{v} is almost aligned with the O-O direction (see inset of Figure \ref{figure3}c), this trend suggests that the hyperfine interaction along the O-O direction is significantly larger than the other two principal axes.

The extensive hyperfine spectra presented above allow us to quantitatively extract the hyperfine interactions along the three principal axes of bridge-site Ti atoms. Taking both $\mathbf{A}$ and $\mathbf{g}$-factor anisotropies under consideration, the experimentally observed hyperfine splitting $A(\theta)$ is related to its principal values by \cite{abragam2012electron} 

% \begin{equation}
% A(\theta) = \sqrt{(A_{O-O}\cos\varphi)^{2}+(A_{Mg-Mg}\sin\varphi)^{2}+ A_{zz}^{2}}.
%     \label{eq: binding site hyperfine}
% \end{equation}

\begin{equation}
g(\theta) A(\theta) = \sqrt{l^{2}g_{\mathrm{O}}^{2}A_{\mathrm{O}}^{2}+m^{2}g_{\mathrm{Mg}}^{2}A_{\mathrm{Mg}}^{2}+n^{2}g_{z}^{2}A_{z}^{2}},
    \label{eq: binding site hyperfine}
\end{equation}
where $(l,m,n)$ are the direction cosines given by $(l,m,n)=(\sin\theta\cos\varphi,\sin\theta\sin\varphi,\cos\theta)$ and $g(\theta)=(l^{2}g_{\mathrm{O}}^{2}+m^{2}g_{\mathrm{Mg}}^{2}+n^{2}g_{z}^{2})^{1/2}$. In our setup, $\varphi_{\mathrm{h}}=15.5^{\circ}$ and $\varphi_{\mathrm{v}}=105.5^{\circ}$ for \textsuperscript{47}Ti\textsubscript{v} and \textsuperscript{47}Ti\textsubscript{h}, respectively. Using the $g$-factors that we measured before on bridge-site Ti, $(g_{\mathrm{O}},g_{\mathrm{Mg}},g_{z})=(1.653,1.917,1.989)$ (Figure \ref{figure3}e) \cite{PhysRevB.104.174408}, the best fits to the hyperfine splittings (solid curves Figure \ref{figure3}f) yield three principal values of hyperfine interaction tensor, $(A_{\mathrm{O}},A_{\mathrm{Mg}},A_{z})=(68.97\pm 1.23,11.66\pm 3.59,21.23\pm 1.52)$. These results indicate a significantly larger hyperfine interaction along the O-O axis compared to the other two axes (Mg-Mg and out-of-plane directions), agreeing with the trends in Figure \ref{figure3}f. The $g$-factor anisotropy is not important in determining the hyperfine constants, as shown by very similar fitting results assuming an isotropic $g$-factor of 2.003 (dashed curves in Figure \ref{figure3}f). The complete determination of the hyperfine tensor $\mathbf{A}$ allows us to calculate its isotropic and anisotropic components (see eq. \ref{eq: spin-Hamiltonian}). Since the dipolar hyperfine tensor $\mathbf{T}$ is traceless, the isotropic hyperfine interaction can be determined experimentally to be $A_{\mathrm{iso}}=1/3\mathrm{Tr}(\mathbf{A})= 33.95 \pm 1.36$ MHz. The dipolar hyperfine tensor $\mathbf{T}$ is then obtained by subtracting $A_{\mathrm{iso}}$ from $\mathbf{A}$, resulting in $(T_{\mathrm{O}},T_{\mathrm{Mg}},T_{z}) = (35.02 \pm 1.84, -22.29 \pm 3.84, -12.72 \pm 2.04)$ MHz.

\begin{figure}[htbp]
  \includegraphics[width=1 \textwidth ]{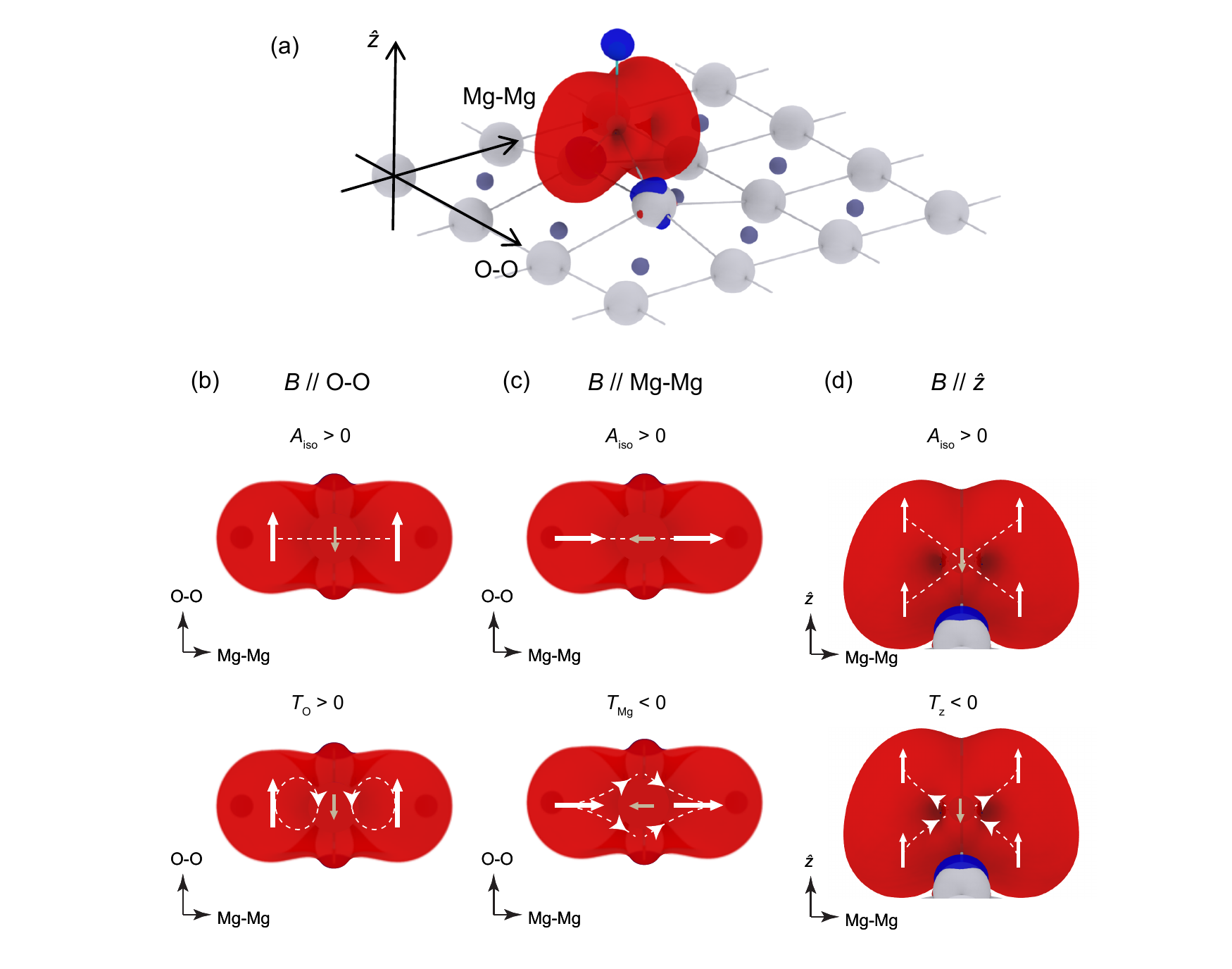}
  \caption{DFT calculations of the spin distribution and hyperfine interactions of bridge-site Ti on MgO. (a) Isometric view of Ti on MgO/Ag with only the top layer of MgO shown for clarity. The red (blue) isosurfaces represent the positive (negative) electron spin polarization (isovalue = 0.002 electrons/$a_0^3$). (b--d) Schematics of isotropic and anisotropic hyperfine interactions with $\boldsymbol{B}_{\mathrm{ext}}$ applied along the three principal axes. The isotropic hyperfine interaction, $A_{\text{iso}}$, is positive and dominates over the dipolar contribution, $T$. As a result, the nuclear magnetic moment (brown arrow) is always anti-aligned with the electron magnetic moment (white arrows) (upper row in (b--d)) (note that the $g$-factors of both the electron and nuclear spins of Ti are negative). The anisotropic dipolar contribution, on the other hand, modifies the hyperfine strength depending on the direction of $\boldsymbol{B}_{\mathrm{ext}}$ (lower row in (b-d)). When $\boldsymbol{B}_{\mathrm{ext}}$ is applied along the O-O direction in (b), a positive dipolar term, $T_\text{O}$, adds to $A_{\text{iso}}$ and is responsible for the largest hyperfine interaction along this principal axis. $\boldsymbol{B}_{\mathrm{ext}}$ applied along the Mg-Mg or $\hat{z}$ direction results in a negative dipolar term that reduces the amplitude of the total hyperfine interaction along these directions.}  
  \label{figure5}
\end{figure}

%\section{DFT and Hyperfine Calculations}
The observed large hyperfine anisotropy along the O-O direction uniquely reflects a highly anisotropic ground-state wavefunction of bridge-site Ti atoms. This is shown by DFT calculations using the GIPAW formalism implemented in Quantum Espresso.\cite{Varini2013,doi:10.1063/5.0005082}. Our model consists of 4 layers of Ag capped by two layers of MgO and a hydrogenated Ti atom adsorbed on a bridge site (Figure \ref{figure5}a and S7). 
%For convenience the coordinate system is rotated so that $\hat{x}$ lies along the O-O and $\hat{y}$ lies along the Mg-Mg in-plane direction of the MgO surface with $\hat{z}$ being the surface normal. 
The resulting DFT ground-state is a mixture of $s$, $d_{yz}$, and $d_{x^2-y^2}$ orbitals, and its spin-polarization isosurface is shown in Figure \ref{figure5}a. We found that the isotropic hyperfine interaction, $A_{\text{iso}}$, is positive and dominates over the dipolar contribution, $T$. As a result, the electron and nuclear spins of Ti are always anti-aligned, and the same holds for their magnetic moments as shown in the upper row of Figure \ref{figure5}b-d (note that the $g$-factors of both the electron and nuclear spins of Ti are negative). The large hyperfine anisotropy along the O-O direction arises from the highly anisotropic ground-state spin distribution (mostly concentrated in the plane spanned by Mg-Mg and $\hat{z}$ directions, see Figure \ref{figure5}a). With the magnetic field applied along the O-O direction, this electron spin distribution results in a positive dipolar term, $T_\text{O} > 0$, which adds to $A_{\text{iso}}$ and is responsible for the largest hyperfine interaction along this principal axis (Figure \ref{figure5}b). A magnetic field applied along the Mg-Mg or out-of-plane direction results in a negative dipolar term that reduces the amplitude of the total hyperfine interaction along these directions (Figure \ref{figure5}c, d). Quantitatively, we note that the GIPAW results overestimate the polarization of the 4$s$ shell resulting in a Fermi contact contribution that is too large ($A_{\textrm{iso}}\approx 170$ MHz). Nevertheless, the correct trend of the hyperfine anisotropy is captured (Figure S7). The difference between our results and previously published DFT works for Ti on thin layers of MgO \cite{Shehada2021,D2CP01224C} stems from the presence of the Ag substrate as well as the hydrogenated state of Ti. Importantly, unlike an earlier experimental study \cite{Willke336} measured with a single-axis magnetic field, the combined use of vector-field ESR spectroscopy and atom manipulation here allows precise characterization of the full hyperfine tensor and hence an accurate description of the ground-state spin distribution. The experimental measurement of the hyperfine tensor, together with $g$-factor anisotropy, yields a remarkable precision in the determination of the electronic ground state of a single paramagnetic center.

In this work, we show how angle-dependent single-atom ESR-STM spectra enable the precise determination of the hyperfine anisotropy and hence ground-state electron distribution. By measuring at the nano-electronvolt energy resolution, the ESR-STM hyperfine spectroscopy adds a unique probe of local electronic structure to the toolbox of STM. We envision that various types of single spin centers can be probed in a similar fashion \cite{RevModPhys.85.961}, enabling the optimization of their quantum spin properties for future applications in spin-based quantum computing and quantum sensing.

While writing this manuscript, we became aware of a similar experiment performed in another group\cite{arXivLaetitia}. Overall, their results agree very well with those presented here.

%In this work we have demonstrated how angle-dependent hyperfine spectroscopy of atoms adsorbed on thin insulating layers can be used to unequivocally determine the electronic ground state. This highlights the power of ESR-STM, which not only can resolve the adsorption site with sub-atomic precision but also allows to determine g-factor and hyperfine tensor thanks to its excellent energy resolution. A combination of these unique features will be critical to determine the ground state of surface spins, enabling optimization of their quantum spin properties for future application in spin qubits.\\

\begin{acknowledgement}

%Please use ``The authors thank \ldots'' rather than ``The
%authors would like to thank \ldots''.

%The author thanks Mats Dahlgren for version one of \textsf{achemso},
%and Donald Arseneau for the code taken from \textsf{cite} to move
%citations after punctuation. Many users have provided feedback on the
%class, which is reflected in all of the different demonstrations
%shown in this document.

This work was supported by the Institute for Basic Science (Grant No. IBS-R027-D1). 

\end{acknowledgement}

\begin{suppinfo}

Supporting experimental data and simulations are provided in the supplementary information.

\end{suppinfo}

\bibliography{220405_ref}

\end{document}